\documentclass[preprints,article,accept,pdftex,moreauthors]{Definitions/mdpi} 
\firstpage{1} 
\pubvolume{1}
\issuenum{1}
\articlenumber{0}
\pubyear{2022}
\copyrightyear{2022}
\datereceived{} 
\dateaccepted{} 
\datepublished{} 
\hreflink{https://doi.org/} 

\usepackage{amsmath}
\usepackage{amssymb}
\usepackage{amstext}
\usepackage{bm}
\usepackage{xcolor}
\usepackage{helvet}
\usepackage{graphicx} 
\usepackage{physics}
\usepackage{qcircuit}

\Title{Generation of Pseudo-Random Quantum States on Actual Quantum Processors}

\TitleCitation{Generation of Pseudo-Random Quantum States on Actual Quantum Processors}

\Author{Gabriele Cenedese $^{1,2}$\orcidD{}, Maria Bondani $^{3}$\orcidC{}, Dario Rosa$^{4,5}$\orcidA{} and Giuliano Benenti $^{1,2,6,*}$\orcidB{}}

\AuthorNames{Gabriele Cenedese, Maria Bondani, Dario Rosa and Giuliano Benenti}

\AuthorCitation{Cenedese, G.; Bondani, M.; Rosa, D.; Benenti, G.}

\address{%
$^{1}$ \quad Center for Nonlinear and Complex Systems, Dipartimento di Scienza e Alta Tecnologia, Universit\`a degli Studi dell'Insubria, via Valleggio 11, 22100 Como, Italy \\
$^{2}$ \quad Istituto Nazionale di Fisica Nucleare, Sezione di Milano, via Celoria 16, 20133 Milano, Italy\\
$^{3}$ \quad Istituto di Fotonica e Nanotecnologie, Consiglio Nazionale delle Ricerche, via Valleggio 11, 22100 Como, Italy\\
$^{4}$ \quad Center for Theoretical Physics of Complex Systems, Institute for Basic Science (IBS), Daejeon - 34126, Korea\\
$^{5}$ \quad Basic Science Program, Korea University of Science and Technology (UST), Daejeon - 34113, Korea\\
$^{6}$ \quad NEST, Istituto Nanoscienze-CNR, I-56126 Pisa, Italy\\
}

\corres{Correspondence: giuliano.benenti@uninsubria.it}

\abstract{The generation of a large amount of entanglement is a necessary condition for a quantum computer to achieve quantum advantage. In this paper, we propose a method to efficiently 
generate pseudo-random quantum states, for which the 
degree of multipartite entanglement is nearly maximal.
We argue that the method is optimal, and use it to benchmark 
actual superconducting (IBM's \textit{ibm\_lagos}) 
and ion trap (IonQ's \textit{Harmony}) quantum processors. 
Despite the fact that \textit{ibm\_lagos} has lower
single- and two-qubit error rates, the overall 
performance of \textit{Harmony} is better thanks 
to low error rate in state preparation and 
measurement and to the all-to-all connectivity of qubits.
Our result highlights the relevance of the qubits network architecture 
to generate highly entangled states.}
\keyword{Quantum computing; NISQ devices; random quantum circuits}

\begin{document}


\section{Introduction}

\label{sec:intro}

Quantum computers working with about 50-100 qubits could perform certain tasks beyond the capabilities of current classical supercomputers~\cite{qcbook,Preskill2018}, 
and quantum advantage for particular problems has been recently claimed~\cite{Martinis2019,Pan2020,Zoller2022}, 
although later simulations on classical supercomputers~\cite{Liu2021,Brumer2022} have 
almost cloased the quantum advantage gap.  
As a general remark, quantum advantage can only be achieved if the precision of the quantum gates is sufficiently high and the executed quantum algorithm generates a sufficiently large amount of entanglement that can overcome classical tensor network methods~\cite{Waintal2020}. 
Therefore, for quantum algorithms, 
multipartite (many-qubit) entanglement is the key resource to achieve an exponential
speed-up over classical computation.
Unfortunately, existing Noisy Intermediate-Scale Quantum (NISQ) devices suffer from 
various noise sources such as noisy gates, coherent errors, and interactions with an uncontrolled environment. Noise limits the size of quantum circuits that can be reliably executed,
so achieving quantum advantage in complex and practically relevant problems is still a formidable challenge. 
It is therefore important to benchmark the progress of currently available quantum computers~\cite{Gambetta19,Benenti21}.

Quantifying entanglement is a demanding task~\cite{Plenio2007,Horodecki2009}. 
In particular, the characterization of multipartite entanglement is not a simple matter, 
since, as the number of subsystems increases, we should  
consider all possible non local correlations among parties in order 
to obtain a complete description of entanglement. Moreover, tomographic state reconstruction 
requires a number of measures that grows exponentially with the number of qubits~\cite{nielsen00}. 
Finally, there is no unique way to characterize multipartite entanglement~\cite{Horodecki2009}.

On the other hand, bipartite entanglement can be probed by means of entanglement entropies.
In particular, we can consider 
the \textit{second order Rényi entropy} of the reduced density matrix for any of the subsystems.
If it is larger than the entropy of the entire system, we can conclude that bipartite 
entanglement exists between the two subsystems. If the overall state is pure, 
the second order Rényi entropy is directly a measure of bipartite entanglement. 
In that case, in order to quantify the amount of multipartite entanglement one can look at the distribution of the Rényi entropy of a subsystem over all possible bipartitions of the total system.
For example Facchi et al. proposed~\cite{Facchi2006} a method based on the probability density of bipartite entanglement between two parts of the total system; one expects that multipartite entanglement will be large when bipartite entanglement is large and does not depend on the bipartition, namely when its probability density is a narrow function centered at a large value.

Computing entanglement entropies requires the knowledge of the density matrix of the system. Unfortunately, probing the density matrix is also a challenging problem, especially as the dimension of the system increases. For this reason it is necessary to indirectly estimate the entropy, for instance using the method proposed by Brydges et al.~\cite{Brydges2019} via randomized measurements.

For random pure quantum states the entanglement content is almost maximal and the purity (and so the second order Rényi entropy) probability distributions is well known. 
Unlike simpler states like W and GHZ, for which the entanglement content is 
essentially independent of the dimension of the system, for random states 
the average multipartite entanglement is an extensive quantity. 
Moreover, random states are relevant in the study of the complexity of quantum 
circuits~\cite{brandao2021} and black holes~\cite{Hayden2007} and 
for benchmarking quantum hardware~\cite{Gambetta19, choi2023}.

The purpose of this paper is to investigate strategies to
efficiently generate highly entangled states and then find a way to quantify the actual amount of entanglement achieved in state-of-the-art quantum hardware.
In particular, we propose a method (hereafter nicknamed \emph{Direct method})
to efficiently generate pseudo-random quantum states for $n$ qubits, 
approximating true random states to the desired accuracy, by means of layers 
where a random 
permutation of the qubits is followed by two-qubit random state generation.
We show that this method converges to true $n$-qubit 
random states by increasing the number of layers
as fast as the circuit implementing two-qubit random unitary gates using the KAK
parametrization of SU(4) (\emph{KAK method})~\cite{vatan2004optimal}, but with a reduced cost in terms of number of CNOT gates. 
We also argue that the proposed method is optimal for pseudo-random quantum state generation. 
Finally, we implement the method to benchmark actual quantum processors. In particular, two different realizations of quantum hardware are compared: 
IBM's superconducting based devices and IonQ's trapped ions based devices.
We show that, despite the fact that superconducting devices have smaller 
error rates than IonQ for one- and two-qubit 
gates, the overall performance is better in trapped ion devices. 
This is mainly due to the complete connectivity of these machines, which allows avoiding 
noisy SWAP gates to implement qubit permutations. 
Our results highlight the importance of quantum hardware architecture in the implementation of quantum algorithms.

The paper is organized as follows. In Sec.~\ref{sec:generation} we discuss 
and compare methods for the generation of pseudo-random states. 
In Sec.~\ref{sec:hardware} we apply the direct method in real quantum hardware, and 
compare the results for IBMQ and IonQ devices, with the 
second order Rényi entropy estimated via the method of Ref.~\cite{Brydges2019}.
Finally, our conclusions are drawn in Sec.~\ref{sec:conc}.

\section{Generation of pseudo-random quantum states}
\label{sec:generation}

In this section we briefly discuss methods of generating pseudo-random states, starting with the exact strategy and ending with our proposal, which will be verified numerically by comparison with the standard KAK method.

Let $\ket{\psi}$ be a pure state that belongs to the Hilbert space $\mathcal{H}=\mathcal{H}_A \otimes \mathcal{H}_B$, where $\mathcal{H}_A$ and $\mathcal{H}_B$ are spanned respectively by $\{\ket{i_A}\}_{1\leq i_A \leq N_A}$ and $\{\ket{i_B}\}_{1\leq i_B \leq N_B}$. $A$ and $B$ are two bipartitions of the 
entire system.
Assuming, without loss of generality, $N_A\leq N_B$, the state admits a Schmidt decomposition~\cite{qcbook}: 
\begin{linenomath}
\begin{equation}
	\ket{\psi}=\sum_{i=1}^{N_A} \sqrt{x_i} \ket{a_i} \otimes \ket{b_i},
\end{equation}
\end{linenomath}
where 
$\{\ket{a_i}\}$ and $\{\ket{b_i}\}$ are a suitable basis sets for 
$\mathcal{H}_A$ and $\mathcal{H}_B$,
which depend on the particular state $\ket{\psi}$, and
the scalars $x_i$, known as the Schmidt coefficients for $\ket{\psi}$, are real, 
non-negative, and unique up to reordering.
These coefficients can be used to quantify 
the bipartite entanglement via the second order Rényi entropy
\begin{equation}
S^{(2)}(\rho_A)=-\log_2 [R(\psi)],
\end{equation}
with the reduced purity $R(\psi)$ of the state given by
\begin{linenomath}
\begin{equation}
	R(\psi)=\text{Tr}(\rho_A^2)=\sum_{i=1}^{N_A} x_i^2,
\end{equation}
\end{linenomath}
where $\rho_A$ is the reduced density matrix (with respect to $\mathcal{H}_B$)
of the overall state $\rho$: 
\begin{linenomath}
\begin{equation}
	\rho_A=\text{Tr}_B(\rho)=\text{Tr}_B(\ket{\psi}\bra{\psi}).
\end{equation}
\end{linenomath}
Hereafter we shall focus on the purity, which is trivially related to the second order Rényi entropy.

In the case of a random state, the 
cumulants of the purities' probability distributions can be calculated exactly \cite{Giraud2007, Lloyd88}, see for more details App.~\ref{sec:RSmoments}.  
In particular, the first cumulants are given by 
\begin{linenomath}
\begin{equation}
 \mu_{N_AN_B}\equiv \langle  R\rangle=\frac{N_A+N_B}{1+ N_A N_B},
 \label{eq:meanpurities}
\end{equation}
\end{linenomath}
\begin{linenomath}
\begin{equation}
 \sigma^2_{N_AN_B}\equiv\langle ( R-\langle  R\rangle)\rangle^2=\frac{2(N_A^2-1)(N_B^2-1)}{(1+ N_AN_B)^2(2+N_AN_B)(3+N_AN_B)},
  \label{eq:meanvariances}
\end{equation}
\end{linenomath}
and they will be used later to verify the quality of random state generation.

In order to generate a true $n$-qubit random state, the ideal (and only) rigorous way would be to apply a random unitary operator, with respect to the Haar measure of the unitary group $SU(N=2^n)$ (neglecting the global phase of no physical significance). 
Unfortunately, the implementation of such operator acting on the $n$-qubit Hilbert space requires a number of elementary quantum gates that is exponential in the number of qubits~\cite{qcbook}. 
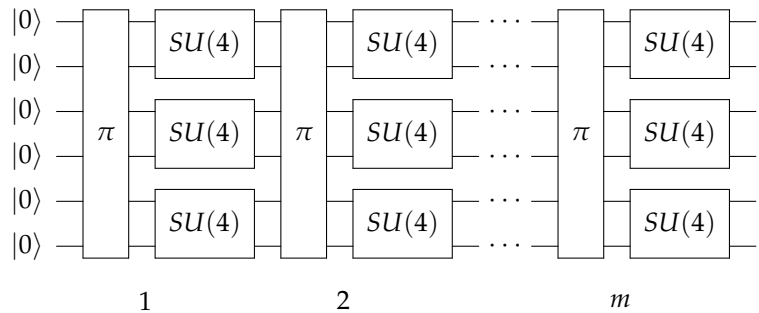
\begin{figure}[H]
	\[
	\begin{array}{c}
		\Qcircuit @C=1em @R=.8em  {
			&\lstick{\ket{0}}&\multigate{5}{\pi}&\multigate{1}{SU(4)}&\multigate{5}{\pi}&\multigate{1}{SU(4)}&\qw&\cdots&&\multigate{5}{\pi}&\multigate{1}{SU(4)}&\qw\\
			&\lstick{\ket{0}}&\ghost{\pi}& \ghost{SU(4)}&\ghost{\pi}& \ghost{SU(4)}&\qw&\cdots&&\ghost{\pi}& \ghost{SU(4)}&\qw\\
			&\lstick{\ket{0}}&\ghost{\pi}& \multigate{1}{SU(4)}&\ghost{\pi}& \multigate{1}{SU(4)}&\qw&\cdots&&\ghost{\pi}& \multigate{1}{SU(4)}&\qw\\
			&\lstick{\ket{0}}&\ghost{\pi}& \ghost{SU(4)}&\ghost{\pi}& \ghost{SU(4)}&\qw&\cdots&&\ghost{\pi}& \ghost{SU(4)}&\qw\\
			&\lstick{\ket{0}}&\ghost{\pi}& \multigate{1}{SU(4)}&\ghost{\pi}& \multigate{1}{SU(4)}&\qw&\cdots&&\ghost{\pi}& \multigate{1}{SU(4)}&\qw\\
			&\lstick{\ket{0}}&\ghost{\pi}&\ghost{SU(4)}&\ghost{\pi}&\ghost{SU(4)}&\qw&\cdots&&\ghost{\pi}&\ghost{SU(4)}&\qw\\	
			&&&&&&&&&&&\\		
		    & &\hspace{3em} 1& & \hspace{3em} 2&&&&&\hspace{3em} m&&}
	\end{array}
	\]
	\caption{The pseudo-random state generator circuit consists of $m$ layers of random permutations of the qubit labels, followed by random two-qubit gates. When the circuit width $n$ is odd, one of the qubits is idle in each layer. In this figure a circuit with $n = 6$ qubits width is shown for illustration purposes.}  
	\label{fig:pseudoRandom}
\end{figure} 
On the other hand, it has been proved that sequences of random single qubit gates followed by a two-qubit local interaction (that can be an $SU(4)$ random unitary operator, or more simply a single CNOT gate) generate pseudo-random unitary operators which approximate, to the desired accuracy, the entanglement properties of true $n$-qubit random states \cite{Emerson03,Emerson2005,Weinstein2005,Dahlsten07,Oliveira07}. However, the random $SU(4)$ strategy depicted in Figure~\ref{fig:pseudoRandom}, used for example in \cite{Cross19},  performs better than a single CNOT in terms of convergence rate \cite{Znidaric07}, with the cost of using 3 CNOTs instead of just one, as we will see below.


Hence, the problem now turns to find an efficient way (in sense that will be clarified later) to generate random $SU(4)$ operators.

To this end, one possible strategy would be to use the Hurwitz's parametrization of the unitary group, $SU(N)$, for the specific case of $N = 4$~\cite{Pozniak98,Weinstein2005}.
However, this approach has the disadvantage of requiring a large number of CNOTs --- $16$ for the particular case of $SU(4)$ --- which are usually the main source of errors in NISQ devices~\cite{Pelofske2022}.

\subsection{Cartan's KAK decomposition of the unitary group}

Alternative approach consists in using the Cartan's KAK decomposition of a semi-simple Lie group $\textbf{G}$ (in this case $SU(2^n)$) which parametrizes the group in terms of subgroups' elements~\cite{humphreys2012}. 
The case of $SU(4)$ of interest here is described in App.~\ref{sec:KAK}, and it is the optimal 
construction~\cite{vatan2004optimal} to implement a generic two-qubit gate, using 
at most 3 CNOT and 15 single-qubit gates.

\subsection{Direct generation of two-qubit random quantum states}
Is the Cartan decomposition the most efficient way to generate a two-qubit random state? 
Let us think in terms of free parameters. 
The Cartan's KAK  decomposition is the optimal 
(in terms of number of CNOT and single-qubit rotations) way to construct random $SU(4)$ operators via quantum circuits. It requires 15 single-qubit rotations and so 15 independent real parameters (as one expects, since $\text{dim}(SU(4))=15$). On the other hand,
a normalized random two-qubit state $\ket{\psi}$ depends, up to a global phase, on 6 independent real parameters. This suggests that in some ways it could be possible to build any two-qubit state (starting from some fiducial state) with at most 6 independent rotations and 1 CNOT (needed to entangle the system).

\begin{figure}[H]
	\[
	\begin{array}{c}
		\Qcircuit @C=1em @R=.8em { 
			&\lstick{\ket{0}}&\multigate{1}{D}&\qw&&&&\lstick{\ket{0}}&\gate{R_y(\theta_1)}&\ctrl{1}&\gate{R_{z}(\theta_2)}&\gate{R_y(\theta_3)}&\gate{R_z(\theta_4)}&\qw\\
			&\lstick{\ket{0}}&\ghost{D}& \qw&&&&\lstick{\ket{0}}&\qw &\targ&\qw&\gate{R_y(\theta_5)}& \gate{R_z(\theta_6)}&\qw 
			\inputgrouph{1}{2}{1.3em}{=}{-5.9em}}
	\end{array}
	\]
	\caption{A circuit for two-qubit random state generation.
 Rotations $R_k$
are obtained by exponentiating the corresponding Pauli matrices $\sigma_k$.}  
	\label{fig:Direct}
\end{figure}
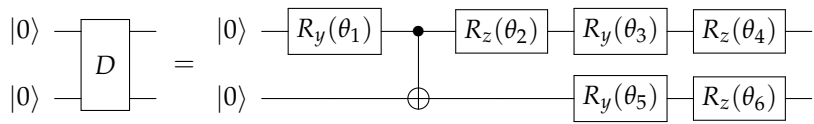

This expectation is confirmed~\cite{Giraud2007} by the quantum
circuit depicted in Fig. \ref{fig:Direct}, producing $\ket{\psi}$ from an initial state $\ket{00}$. 
How can this circuit be achieved? Starting from a state $\ket{\psi}$, and transforming it by the inverse of the circuit of Fig. \ref{fig:Direct} one can end up with $\ket{00}$, specifying how the angles $\theta$ are obtained.
Any two-qubit state in fact can be written, using the Schmidt decomposition, as a sum of two product terms:
\begin{linenomath}
\begin{equation}
	\ket{\psi}=\sqrt{x_1}\ket{a}\ket{b}+\sqrt{x_2}\ket{a}^\perp\ket{b}^\perp,
\end{equation}
\end{linenomath}
where $\ket{a}$ and $\ket{b}$ are single-qubit states (of the first and second qubit, respectively), $\ket{a}^\perp$ and $\ket{b}^\perp$ are single-qubit states orthogonal to $\ket{a}$ and $\ket{b}$ (i.e. $\{ \ket{a},\ket{a}^\perp \}$ and $\{ \ket{b},\ket{b}^\perp \}$ are the Schmidt bases of the Hilbert spaces of the two qubits).
The idea is, starting from this decomposition, to get the state $\ket{00}$ using unitary operations, and then, taking the inverse transformation, one can get the desired result. The angle $\theta_4$ is chosen such that the $R_z$ rotation of angle $-\theta_4$ eliminates a relative phase between the coefficients of the expansion of $\ket{a}$ into $\ket{0}$ and $\ket{1}$ (note that because the circuit is considered in the reverse direction, the angles of rotations have opposite signs). A subsequent $R_y$ rotation with angle $-\theta_3$ results in the transformation $\ket{a}\rightarrow\ket{0}$ (up to a global phase). Similarly, rotations of angles $-\theta_6$ and $-\theta_5$ rotate $\ket{b}$ into $\ket{0}$. After applying rotations of angles $-\theta_3$, $-\theta_4$, $-\theta_5$ and $-\theta_6$ the state has become, up to a global phase, of the form:
\begin{equation}
	\ket{\psi}=\cos\theta_1\ket{00}+e^{i\theta_2} \sin\theta_1\ket{11}.
\end{equation}
Finally the $R_z$ rotation of $-\theta_2$ eliminates the relative phase between $\ket{00}$ and $\ket{11}$. The CNOT brings the second qubit to $\ket{0}$, and the last rotation of angle $\theta_1$ on the first qubit yields the final state $\ket{00}$.

In order to obtain a random state it is necessary to know how to randomly sample the various angles $\theta_i$, that is, it is necessary to know their probability distributions with respect to some measure of the state, associated with the parametrization provided by Fig. \ref{fig:Direct}.
Formally, a quantum state $\ket{\psi}$ can be considered as an element of the complex projective space $\mathbb{C}\text{\textbf{P}}^{N-1}$, with $N=2^n$ the Hilbert space dimension for n qubits~\cite{bengtsson_życzkowski_2017}. 
The natural Riemannian metric on $\mathbb{C}\text{\textbf{P}}^{N-1}$ is the Fubini-Study metric, induced by the unitarily invariant Haar measure on $U(N)$. Thus is the only metric invariant under unitary transformations. 
 Thus, the unnormalized joint probability distribution is simply obtained by calculating the determinant of the metric tensor with the parametrization~\cite{Giraud2009}. 
The idea is to use these more efficient "operators" $D$ to construct the $n$-qubit pseudo-random states, although formally they do not map the entire Bloch sphere if the initial state is not $\ket{00}$. 
Consider, for example, a dimensionally simpler case: from the north pole of a sphere it is possible to reach any other point by making only two rotations. Thus, carefully choosing the distribution of the rotation angles it is possible to uniformly map every point of the sphere, but this is no longer valid if the starting point is changed, where the worst case scenario is a point on the equator.

In general one expects that the error committed in sampling the Bloch space is small and everything converges to a random state anyway
(see below). In Fig.~\ref{fig:pseudoRandomD} it is shown the circuit used to generate the random state (in a similar way to Fig.~\ref{fig:pseudoRandom})  with this method, which will be referred to from now on as the Direct method.

\begin{figure}[h!]
	\[
	\begin{array}{c}
		\Qcircuit @C=1em @R=.8em  {
			&\lstick{\ket{0}}&\multigate{5}{\pi}&\multigate{1}{D}&\multigate{5}{\pi}&\multigate{1}{D}&\qw&\cdots&&\multigate{5}{\pi}&\multigate{1}{D}&\qw\\
			&\lstick{\ket{0}}&\ghost{\pi}& \ghost{D}&\ghost{\pi}& \ghost{D}&\qw&\cdots&&\ghost{\pi}& \ghost{D}&\qw\\
			&\lstick{\ket{0}}&\ghost{\pi}& \multigate{1}{D}&\ghost{\pi}& \multigate{1}{D}&\qw&\cdots&&\ghost{\pi}& \multigate{1}{D}&\qw\\
			&\lstick{\ket{0}}&\ghost{\pi}& \ghost{D}&\ghost{\pi}& \ghost{D}&\qw&\cdots&&\ghost{\pi}& \ghost{D}&\qw\\
			&\lstick{\ket{0}}&\ghost{\pi}& \multigate{1}{D}&\ghost{\pi}& \multigate{1}{D}&\qw&\cdots&&\ghost{\pi}& \multigate{1}{D}&\qw\\
			&\lstick{\ket{0}}&\ghost{\pi}&\ghost{D}&\ghost{\pi}&\ghost{D}&\qw&\cdots&&\ghost{\pi}&\ghost{D}&\qw\\	
			&&&&&&&&&&&\\		
			& &\hspace{3em} 1& & \hspace{3em} 2&&&&&\hspace{3em} m&&}
	\end{array}
	\]
	\caption{The pseudo-random state generator circuit consists of $m$ layers of random permutations of the qubit labels, followed by random $D$ gates. In this figure a circuit with $n = 6$ qubit's width is shown for illustrative purposes.}  
	\label{fig:pseudoRandomD}
\end{figure}
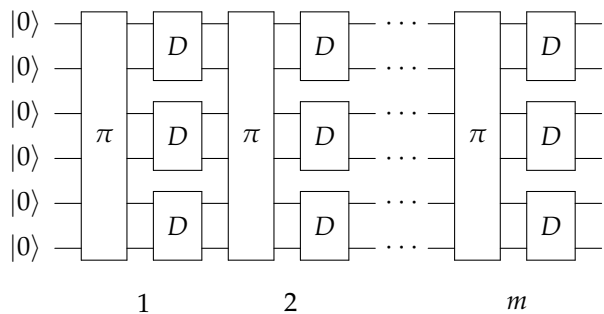 

\subsection{Comparison of KAK and Direct method}

In general, for a $n$-qubit state there are $\binom{n}{n_a}$ ways to construct a bipartition in $n_a$ and $n_b=n-n_a$ qubits
($N_A=2^{n_a}, N_B=2^{n_b}$). Clearly $n_a$ can be any natural number from 1 to $n$. For the sake of clarity let's consider for example a 4-qubit state, the qubits are labeled as $\{ 0,1,2,3\}$. The $n_a=2$ bipartition can be obtained in $\binom{4}{2}=6$ different ways tracing out the pair of qubits $\{0,1\}$, $\{0,2\}$, $\{0,3\}$, $\{1,2\}$, $\{1,3\}$ or $\{2,3\}$. In the case of a random state, these partitions are equivalent, i.e., the value of purity is independent of the choice of the subset of qubits traced out.

Given a quantum state generated as shown in Fig.~\ref{fig:pseudoRandom} (KAK method) and Fig.~\ref{fig:pseudoRandomD} (Direct method), 
and taking an ensemble of $N_{e}$ states, we numerically estimate the mean value $(\mu_{2^{n_a} 2^{n-n_a}})_e$ and the variance $(\sigma^2_{2^{n_a} 2^{n-n_a}})_e$ 
of the purities of the generated pseudo-random quantum state. Simulations are performed using the Python library \textit{Qiskit}, in particular the system density matrix is computed using the built-in state-vector simulator.
In order to evaluate how well the states are generated, the idea is to calculate the relative error of the mean value $\Delta_\mu$ and the variance $\Delta_{\sigma^2}$, which are averaged over each possible bipartition of the number of qubits:
\begin{equation}
	\overline{\Delta_\mu}= \frac{1}{n-1}\sum_{n_a=1}^{n-1} \frac{|(\mu_{2^{n_a} 2^{n-n_a}})_e-\mu_{2^{n_a} 2^{n-n_a}}|}{\mu_{2^{n_a} 2^{n-n_a}}},
\end{equation}
\begin{equation}
	\overline{\Delta_{\sigma^2}}= \frac{1}{n-1}\sum_{n_a=1}^{n-1} \frac{|(\sigma^2_{2^{n_a} 2^{n-n_a}})_e-\sigma^2_{2^{n_a} 2^{n-n_a}}|}{\sigma^2_{2^{n_a} 2^{n-n_a}}}.
\end{equation}
The sum on $n_a$ is up to $n-1$, since $n_a=n$ means tracing out the whole system, i.e. calculating the purity of the whole state, which, being pure, has unit mean and zero variance.
In the previous formulas, the quantities $\mu_{2^{n_a} 2^{n-n_a}}$ and $\sigma^2_{2^{n_a} 2^{n-n_a}}$ are the expected values for a true random state, Eqs.~\eqref{eq:meanpurities} and \eqref{eq:meanvariances}, respectively.

\begin{figure}[H]
	\includegraphics[width=15 cm]{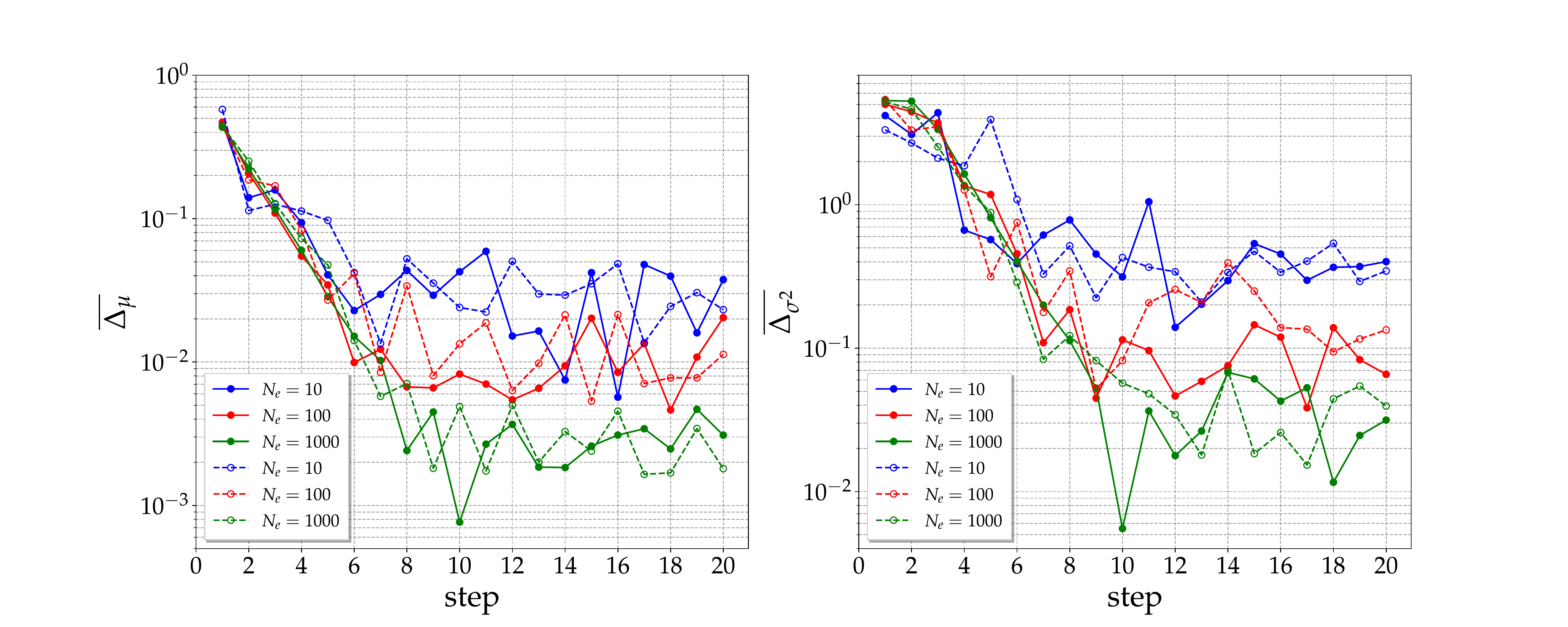}
	\includegraphics[width=15 cm]{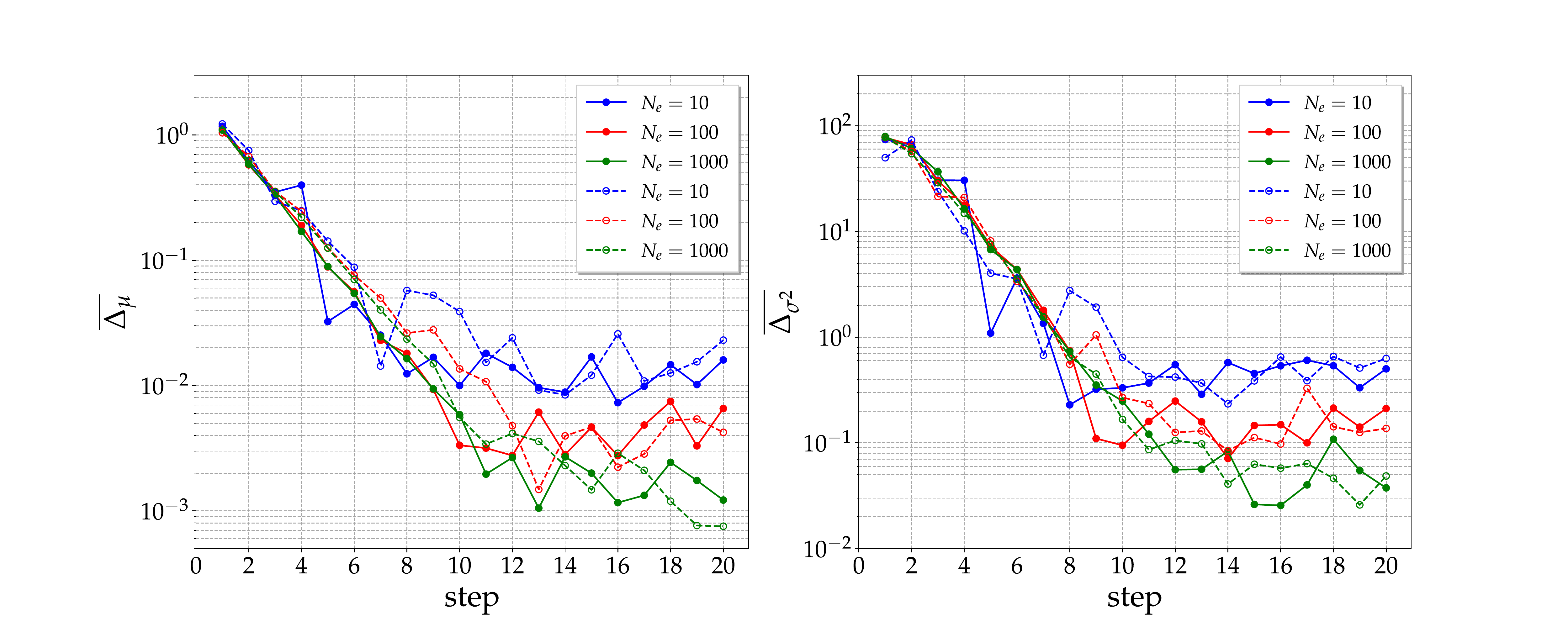}
	\includegraphics[width=15 cm]{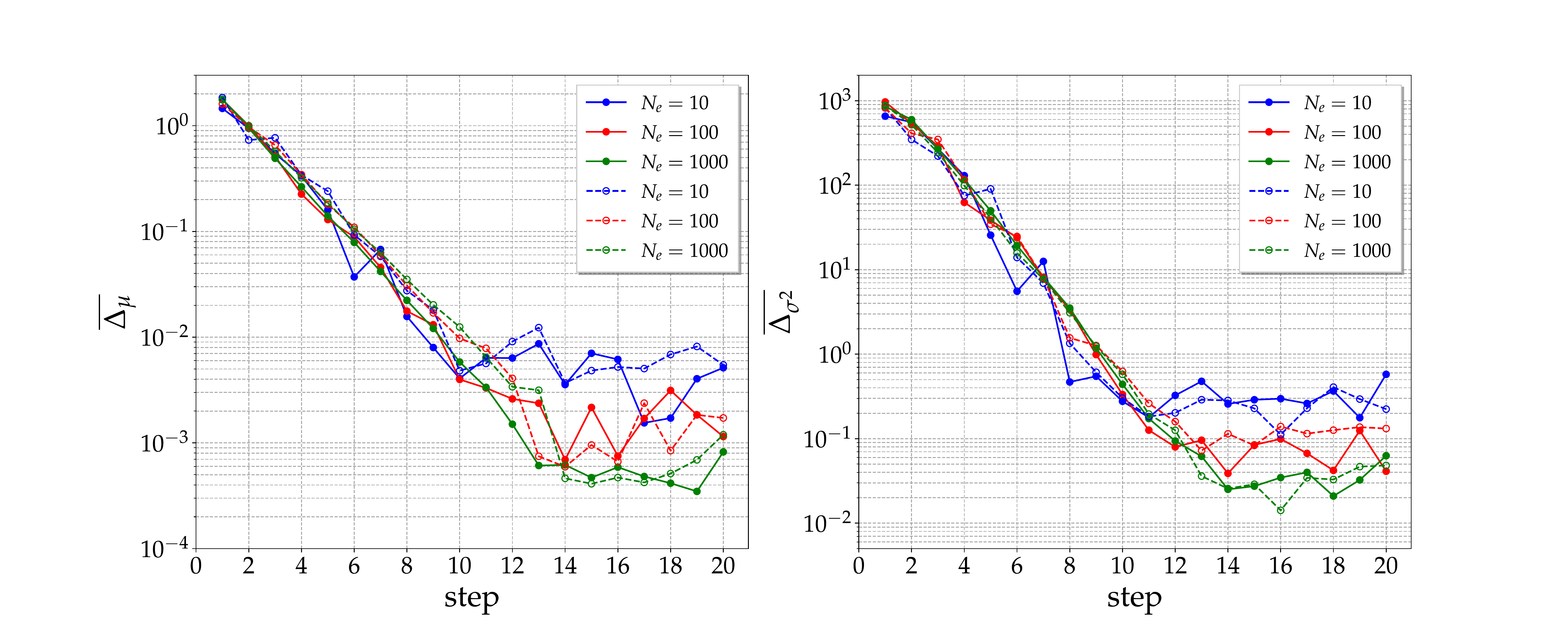}
	\caption{Average mean value relative error (left) and average variance relative error (right) for purities as a function of the number of steps (i.e., layers in the quantum circuit) and the ensemble size for 4-qubit (top), 6-qubit (middle) and 8-qubit (bottom) pseudo-random quantum state. The solid lines represent the Direct method while the dashed lines represent the KAK method.
		\label{fig:directVSkak}}
\end{figure}

The averaged relative error for the mean and variance 
is plotted as a function of the number of steps (i.e., of layers)  of the generating circuit and the size of the statistical ensemble, for the cases of $n=4$,  6, and 8 qubits. 
As it can be seen from Fig.~\ref{fig:directVSkak} the Direct and the KAK methods are basically equivalent. 
Notice that the number of steps required for convergence grows as $\sim n$, since at least $n(n-1)/2$ two-qubit gates are required
in order to tangle all qubit pairs, and for each step $n/2$ two-qubit
gates are applied.

\begin{figure}[H]
	\includegraphics[width=13 cm]{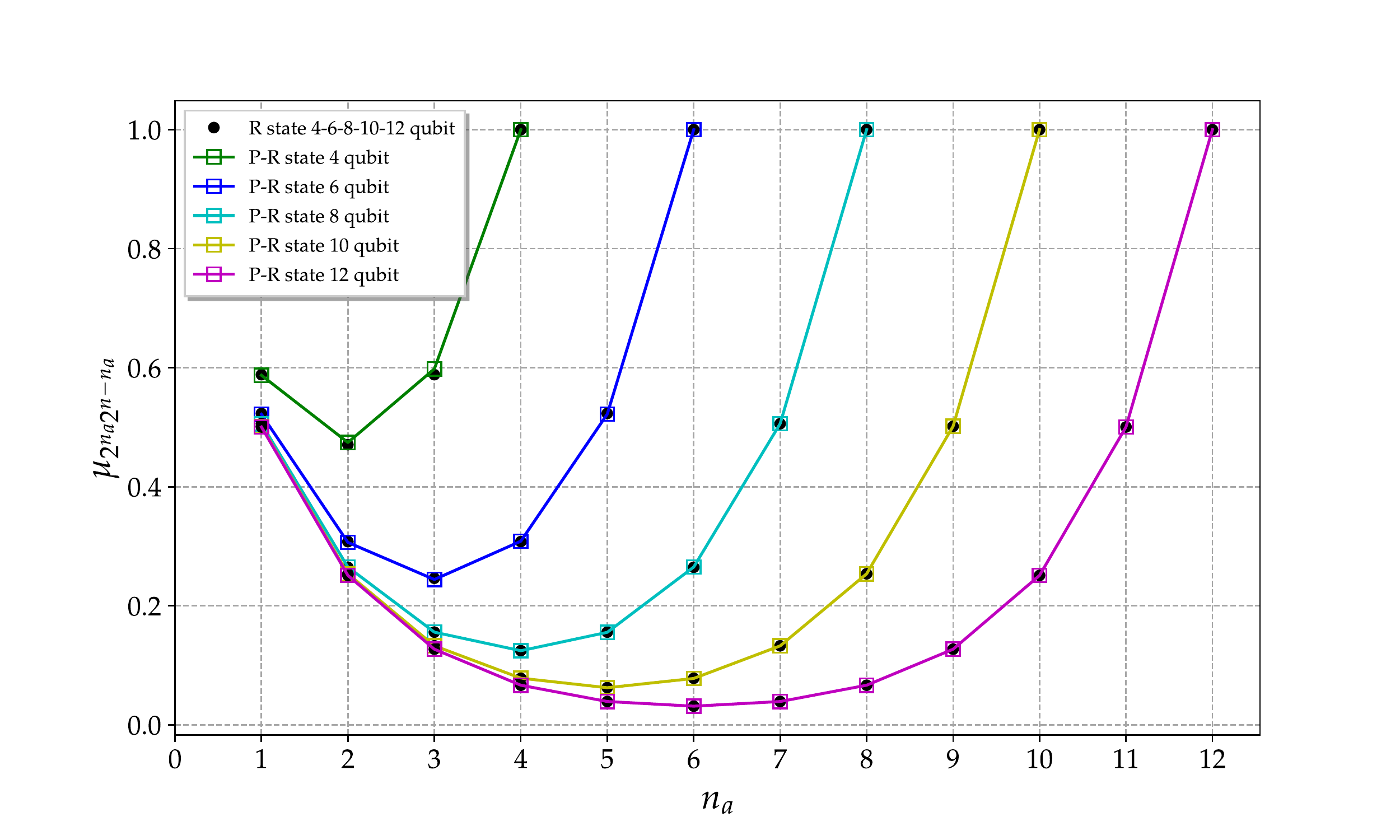}
    \includegraphics[width=13 cm]{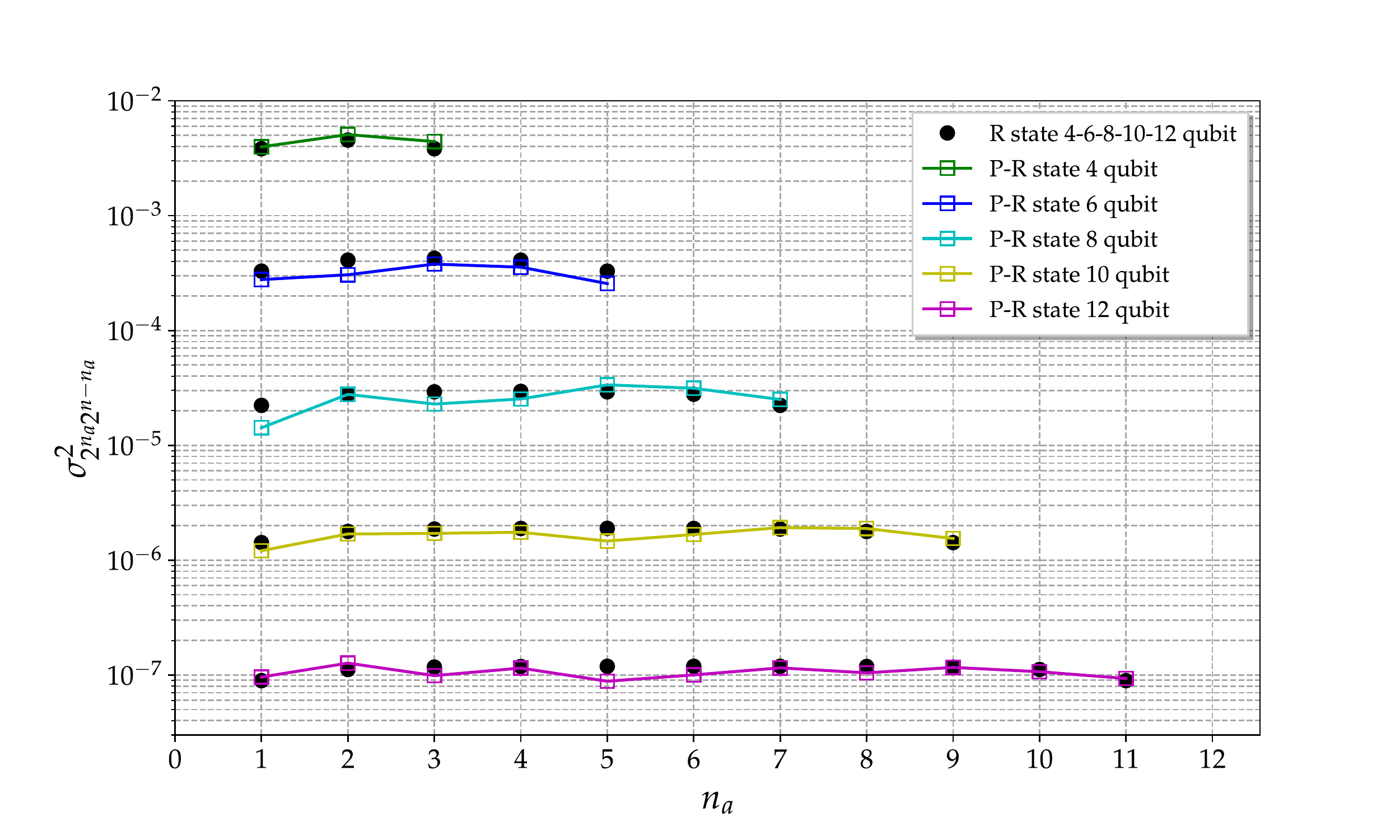}
	\caption{Purity mean value (top) and variance (bottom) of a pseudo-random quantum state plotted as a function of partition size. The various colors represent systems of different dimensions (number of qubits). The black dots are the expected values for a true random state. Here is shown the Direct method with 20 steps and $N_{e}=100$.
	\label{fig:meanNvar}}
\end{figure}

In Fig.~\ref{fig:meanNvar} the mean value and the variance of the purities are shown as a function of the number of qubits in a partition $n_a$, for systems with different size $n$. The moments are estimated considering an ensemble of $N_{e}=100$ pseudo-random states generated using the Direct method with 20 steps. 
The convergence to the true random state expected values improve as the dimension of the system increases. Indeed, for higher dimensions, the entanglement content is highly typical, i.e. it is possible to show that the entanglement distribution for a random state becomes 
strongly peaked in the limit of a large number of qubits. This concentration of the measure  
explains the better convergence for higher dimensional case~\cite{Dahlsten2014}.

\section{Results on actual quantum hardware}

\label{sec:hardware}

\begin{figure}[H]
	\includegraphics[width=7 cm]{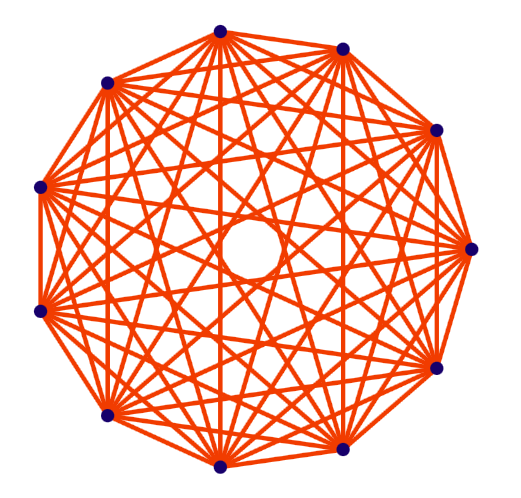}
    \includegraphics[width=7 cm]{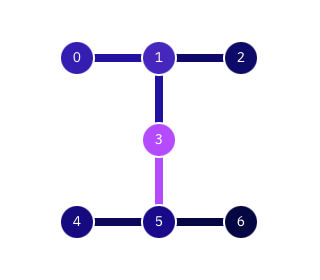}
	\caption{Architectures of the quantum processors used in this work. The circles represent the qubits while the lines represent the physical connection between them. On the left the architecture of IonQ’s $Harmony$, where is clear the complete connectivity of ion based devices. On the  right we have $ibm\_lagos$. Here the color scheme (blue for min, violet for max) refers to the single-qubit (color of the circles) and two-qubit (color of the lines) error rates. These are purely indicative since the rates change upon every calibration of the devices.
	\label{fig:arc}}
\end{figure}

The circuits we have implemented on real quantum hardware 
(IBM's \textit{ibm\_lagos} 
and IonQ's \textit{Harmony}, see for a visualization
Fig.~\ref{fig:arc}) 
are slightly different from that shown in Fig.~\ref{fig:pseudoRandomD}. 
First of all, given the available resources, 
only circuits with 4 and 6 qubits have been considered.
In order to limit circuit depth, the random permutation gates are avoided, and instead, since all the qubits must be tangled with each others, the $D$ (or $SU(4)$) gates are applied to qubit pairs labeled as $\{$(0,1), (2,3), (0,2), (1,3)$\}$ (for the 4 qubits case) and $\{$(0,1), (2,3), (4,5), (1,2), (3,4), (0,5), (0,3), (1,4), (5,2$)\}$ (for the 6 qubits case).
The purities of a random state are estimated using measurements along randomly rotated axes,
following the method proposed by Brydges et al.~\cite{Brydges2019}.

The ensemble of random states is $N_e=10$ wide, and for each state $N_m=20$ random measurement axes are taken in order to estimate the purities. Each of these $200$ circuits is followed by a measurement in the standard computational basis, and each circuit is repeated $N_s=1000$ times (number of shots, limited by the available budget 
for IonQ), in order to estimate the outcome probabilities of each element of the computational basis for each circuit.

Note that IBM’s quantum computers are nominally calibrated once over a 24-hour period, and the system properties update once this calibration sequence is complete. Calibration plays a critical 
role in quantum circuit execution, since the properties of the systems are utilized for noise-aware circuit mapping and optimization (transpilation).
Due to the daily calibration it is difficult to compare results obtained in different days on the same hardware.
For this reason here all comparative results with the same quantum computer were taken with the same calibration data (i.e., the same day).

\subsection{Comparison between hardware platforms}

From extensive tests performed in the literature~\cite{Pelofske2022} (see Table~\ref{tab:QPU}) we know that 
IBM's \textit{ibm\_lagos} has a better performance than IonQ's \textit{Harmony} as far as mean fidelities
for one- and two-qubit gates are considered. On the other hand, IonQ's \textit{Harmony}
is preferable when State Preparation and Measurement (SPAM) fidelities are considered.
More importantly, IonQ's \textit{Harmony} has the advantage of an all-to-all connectivity.
This latter point is very relevant, because IBM's quantum processors need SWAP gates 
to implement  $D$ (or SU(4)) gates between qubits not connected.
Moreover, a SWAP gate is not a native gate on the IBMQ devices, 
and must be decomposed into three CNOT gates. 
Being the product of three CNOT gates, SWAP gates are expensive operations to perform on a 
noisy quantum device.

\begin{table}[H]
	\begin{adjustwidth}{-\extralength}{0cm}
	\newcolumntype{C}{>{\centering\arraybackslash}X}
	\begin{tabularx}{\fulllength}{CCCCCCCCC}
		\toprule
		&&&&\textbf{QPU}&&&\textbf{Fidelity}&\\
		\cmidrule{4-6} 
		\textbf{Vendor} & \textbf{Backend} & \textbf{QV} & \textbf{\# Qubit} & \textbf{Topology} & \textbf{\# Edges}& \textbf{2Q gate}&\textbf{1Q gate} &\textbf{SPAM} \\
		\midrule
		IBM Q& $ibm\_lagos$& 32 &7 & Falcon~r5.11H & 6&0.9924 &0.9998 &0.9862 \\
		IonQ&$Harmony$ &8* &11 &All-to-All & 55&0.96541 & 0.9972&0.99709 \\
		\bottomrule
	\end{tabularx}
\end{adjustwidth}
\caption{Table of Quantum Processing Units (QPUs) evaluated in~\cite{Pelofske2022} using the Quantum Volume (QV) protocol. Values of QV, as well as single-qubit (1Q) gate, two-qubit (2Q) gate and State Preparation and Measurement (SPAM) fidelities are all vendor provided metrics. The mean gate and SPAM fidelities are computed in~\cite{Pelofske2022} across all operations of the same type available on the device during the whole QV circuit execution duration. The number of edges for each backend was counted simply as the number of connections between qubits. \\ * The QV value for IonQ's $Harmony$ is the one measured in~\cite{Pelofske2022}, since IonQ does not provide it.\label{tab:QPU}}
\end{table}

\begin{figure}[H]
	\includegraphics[width=13 cm]{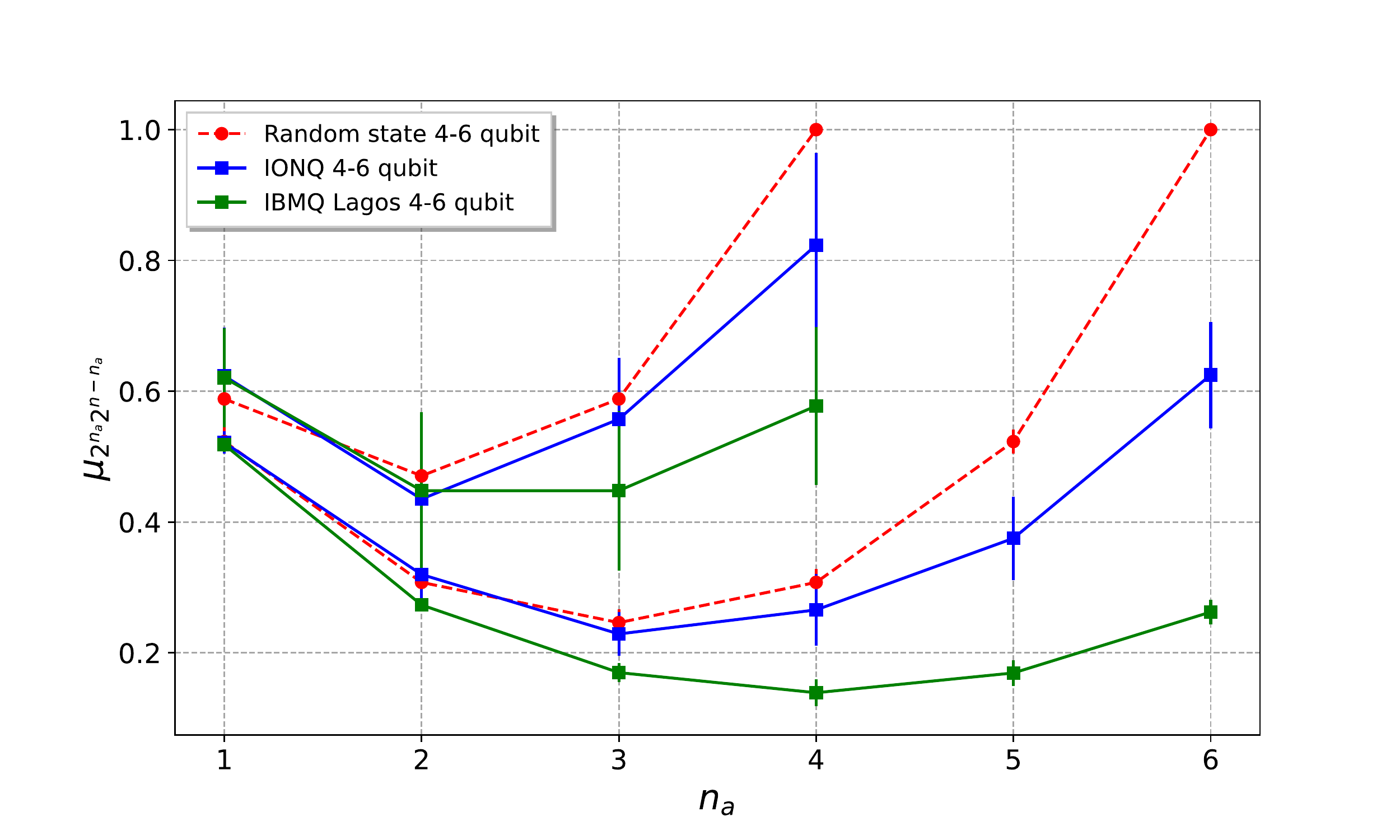}
	\caption{Comparison between the purities of a 4 and 6 qubits pseudo-random quantum state, generated in the two different realizations of a quantum computer investigated, with the Direct method. In green superconductors IBM's \textit{ibm\_lagos} while in blue IonQ's \textit{Harmony}. Red curves give the results for ideal random states. Data were obtained on September 10th, 2022, for \textit{ibm\_lagos} and on July 24th, 2022, for \textit{Harmony}.
	\label{fig:IBMvsIONQ}}
\end{figure}

The results obtained using the Direct method are shown in Fig.~\ref{fig:IBMvsIONQ}, both for IBMQ and IonQ. As it can be seen from the figure, particularly in IonQ case, the purity of the whole state is greater than the bipartitions reduced purities, with the exception of the $n_a=1$ case for the IBMQ. 
This is equivalent to say that the entropies of the parts are greater than the entropy of the whole state, which is a signature of bipartite entanglement in the system. 
Despite the fact that superconductor devices have lower 
error rates than IonQ for single-qubit and two-qubit gates, the overall purity is higher in trapped ion devices. This is mainly due to the complete connectivity of these machines, which allows avoiding noisy SWAP gates, 
in addition to to the better SPAM fidelities of the ion based device.


\label{sec:results}

 \subsection{Entanglement evolution}

\begin{figure}[H]
	\includegraphics[width=14 cm]{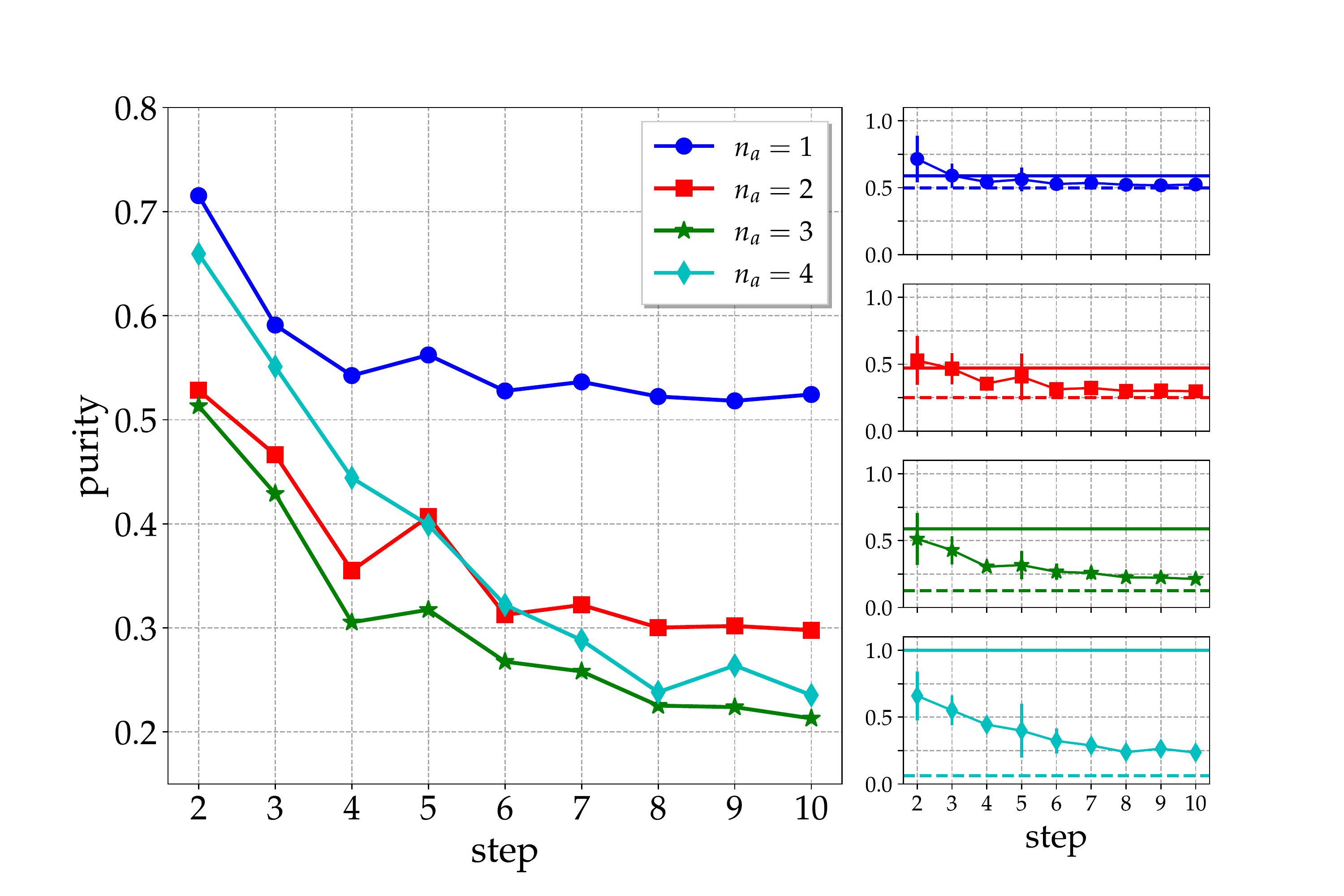}
	\caption{Evolution of the entanglement content of a pseudo-random quantum state generated by the circuit described in Fig.~\ref{fig:pseudoRandomD} as a function of the number of layers (steps). The panels on the right show the individual curves, with the horizontal solid lines highlighting the purity expectation values for a true random state. The horizontal dashed lines refer to the purity of a maximally mixed state. Data taken from $ibm\_lagos$ on January 29, 2023.}
	\label{fig:evo}
\end{figure}

 To investigate the survival of entanglement in an operating quantum computer, we iterate the above circuit for the generation of 
 pseudo-random quantum states for a number of steps.
 In Fig.~\ref{fig:evo} we 
 consider \textit{ibm\_lagos} and $n=4$ qubits, 
 and show the purities as a function 
 of the number of steps, for subgroups of $n_a$ qubits.
 We can see that the purity of the overall system 
 (ideally pure) is clearly higher that the purities of subsystems 
 with $n_a=2$ and $n_a=3$ qubits up to $4$ steps.
For longer evolution times, the purity of the overall 
system drops below those of subsystems, and there
is evidence, at least for $n_a=1,2$, of convergence 
to the purity for a maximally mixed state, 
equal to $1/2^{n_a}$. These values are smaller than 
those for pseudo-random states,
reported in Eq.~(\ref{eq:meanpurities}).
Overall the above remarks point to a vanishing 
entanglement content in the quantum hardware after 
$4-5$ steps. 

\label{sec:evolution}




\section{Conclusions}

\label{sec:conc}

We have investigated the generation of random states, for which the entanglement content is almost maximal, on a quantum computer. 
We have proposed a method in which the obtained pseudo-random states converge to true random states by concatenating layers in which random permutations of the qubit labels are followed by the generation of random states for pairs of qubits. We have argued that our method is optimal,
and the number of CNOT gates is greatly reduced with respect to
circuits implementing two-qubit random unitary gates.
The effectiveness of our method has been tested in current 
implementations of quantum hardware, both for superconducting and ion trap quantum processors. In the latest implementation, we 
have highlighted the advantages of the all-to-all connectivity of qubits.

With regard to the attainment of maximal entanglement of quantum states, it would be interesting to study the class of maximally multipartite $n$-qubit states proposed by Facchi et al.~\cite{Facchi2008}.
More generally, multipartite entanglement optimization is a difficult task, which could at the same time be an ideal testbed for investigating the complexity of quantum correlations in many-body systems and for developing variational hybrid quantum-classical algorithms~\cite{McClean2016,Moll2018,Cerezo2021}. 

\vspace{6pt} 


\authorcontributions{G. C. performed quantum simulations by coding actual IBM quantum processors. G. B. supervised the work with inputs from M. B. and D. R.. All authors discussed the results and contributed to writing and revising the manuscript.}

\funding{G. C. and G. B. acknowlewdges the financial support of the INFN through the project QUANTUM. D.~R. acknowledges the support by the Institute for Basic Science in Korea (IBS-R024-D1).}

\institutionalreview{Not applicable.}

\informedconsent{Not applicable.}

\dataavailability{The dataset used and analyzed in the current study are available from the corresponding author on reasonable request.} 

\acknowledgments{We acknowledge use of the IBM Quantum Experience and the access to IonQ machines supported by Amazon Web Service. The views expressed in this work are those of the authors and do not reflect the official policy or position of IBM and IonQ companies.}

\conflictsofinterest{The authors declare no conflict of interest. The funders had no role in the design of the study; in the collection, analyses, or interpretation of data; in the writing of the manuscript, or in the decision to publish the results.}





\appendixtitles{yes} 
\appendixstart
\appendix

\section[\appendixname~\thesection]{Random state purities moments}
\label{sec:RSmoments}
Recalling that $\ket{\psi}$ is a pure state that belongs to the Hilbert space $\mathcal{H}=\mathcal{H}_A \otimes \mathcal{H}_B$, where $\mathcal{H}_A$ and $\mathcal{H}_B$ are spanned respectively by $\{\ket{i_A}\}_{1\leq i_A \leq N_A}$ and $\{\ket{i_B}\}_{1\leq i_B \leq N_B}$, $A$ and $B$ are two bipartitions of the entire system. The state, assuming $N_A\leq N_B$, admits a Schmidt decomposition~\cite{qcbook}: 
\begin{linenomath}
\begin{equation}
	\ket{\psi}=\sum_{i=1}^{N_A} \sqrt{x_i} \ket{a_i} \otimes \ket{b_i},
\end{equation}
\end{linenomath}
where 
$\{\ket{a_i}\}$ and $\{\ket{b_i}\}$ are a suitable basis sets for 
$\mathcal{H}_A$ and $\mathcal{H}_B$.

For a pure random state the Schmidt coefficients $x_i$ are distributed according to the density \cite{Lloyd88}:
\begin{equation}
	P(x_1,\dots,x_{N_A})=\mathcal{N}\prod_{1\leq i < j \leq N_A}(x_i-x_j)^2 \prod_{1\leq k \leq N_A } x_k^{N_B-N_A} \delta\biggl(1-\sum_{i=1}^{N_A}x_i\biggr),
\end{equation}
for $x_i\in [0,1]$ and some normalization factor $\mathcal{N}$. From this distribution it is possible to calculate the $n$-th moment of the purities, defined as \cite{Giraud2007}: 
\begin{align}
	\langle R^n\rangle\notag&=\mathcal{N}\int_{0}^{1}\text{d}x_1\dots \text{d}x_{N_A} \, (x_1^2+x_2^2+\dots+x_{N_A}^2)^n P(x_1, \dots,x_{N_A})= \\
	&=\frac{(N_AN_B-1)!}{(N_AN_B+2n-1)!} \sum_{n_1+n_2+\dots+n_{N_A}=n}\frac{n!}{n_1!n_2!\dots n_{N_A}!} \times\\
	\notag&\times \prod_{n_i\neq0} \biggl[ \frac{(N_B+2n_i -i)!(N_A+2n_i-i)!}{(N_B-i)!(N_A-i)!(2n_i)!}\prod_{j=1}^{i-1} \biggl( 1-\frac{2n_j}{2n_i+j-i}\biggr) \biggr].
\end{align}
Out of this last formula it is easy to calculate the cumulants shown in~Eqs. \ref{eq:meanpurities} and \ref{eq:meanvariances}.

\section[\appendixname~\thesection]{Cartan's KAK decomposition of the unitary group}
\renewcommand{\thefigure}{B\arabic{figure}}
\setcounter{figure}{0}
\label{sec:KAK}
The Cartan's KAK decomposition can be used for constructing an optimal quantum circuit for achieving a general two-qubit quantum gate, up to a global phase, which requires at most 3 CNOT and 15 elementary one-qubit gates from the family $\{R_y,R_z\}$, i.e. single-qubit rotations obtained by exponentiating the corresponding Pauli matrices. It can be proved that this construction is optimal, in the sense that there is no smaller circuit, using the same family of gates, that achieves this operation \cite{vatan2004optimal}. 
\begin{figure}[H]
    \centering
	\includegraphics[width=11 cm]{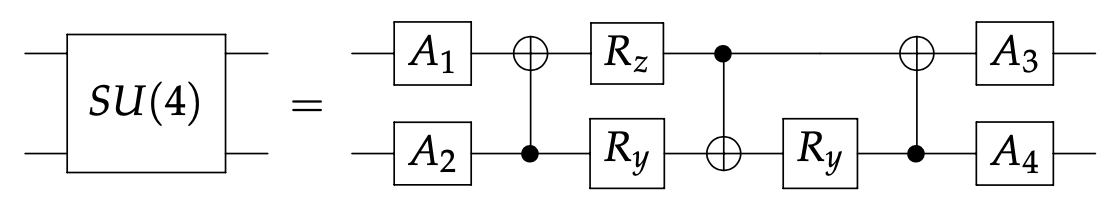}
	\caption{A quantum circuit implementing a two qubits unitary gate using the KAK parametrization of $SU(4)$.}  
	\label{fig:KAK}
\end{figure} 
Following the general prescription \cite{khaneja2001cartan,khaneja2001time}
one can decompose every $SU(4)$ element as depicted in Figure~\ref{fig:KAK}, where $A_j\in SU(2)$ are single qubit unitaries decomposable into elementary one-qubit gates according to the well-known Euler decomposition. Note that in order to randomly extract one of these operators the angles of the single qubit rotations must be extracted uniformly with respect to the Haar measure of the unitary group.

\begin{adjustwidth}{-\extralength}{0cm}

\reftitle{References}



\externalbibliography{yes}
\bibliography{main.bib}

\begin{thebibliography}{999}

\bibitem[Benenti \em{et~al.}(2019)Benenti, Casati, Rossini, and Strini]{qcbook}
Benenti, G.; Casati, G.; Rossini, D.; Strini, G.
\newblock {\em Principles of Quantum Computation and Information (A
  Comprehensive Textbook)}; World Scientific Singapore,  2019.

\bibitem[Preskill(2018)]{Preskill2018}
Preskill, J.
\newblock Quantum {C}omputing in the {NISQ} era and beyond.
\newblock {\em {Quantum}} {\bf 2018}, {\em 2},~79.
\newblock {\url{https://doi.org/10.22331/q-2018-08-06-79}}.

\bibitem[Arute \em{et~al.}(2019)Arute, Arya, Babbush, Bacon, Bardin, Barends,
  Biswas, Boixo, Brandao, Buell, Burkett, Chen, Chen, Chiaro, Collins,
  Courtney, Dunsworth, Farhi, Foxen, Fowler, Gidney, Giustina, Graff, Guerin,
  Habegger, Harrigan, Hartmann, Ho, Hoffmann, Huang, Humble, Isakov, Jeffrey,
  Jiang, Kafri, Kechedzhi, Kelly, Klimov, Knysh, Korotkov, Kostritsa, Landhuis,
  Lindmark, Lucero, Lyakh, Mandr{\`a}, McClean, McEwen, Megrant, Mi,
  Michielsen, Mohseni, Mutus, Naaman, Neeley, Neill, Niu, Ostby, Petukhov,
  Platt, Quintana, Rieffel, Roushan, Rubin, Sank, Satzinger, Smelyanskiy, Sung,
  Trevithick, Vainsencher, Villalonga, White, Yao, Yeh, Zalcman, Neven, and
  Martinis]{Martinis2019}
Arute, F.; Arya, K.; Babbush, R.; Bacon, D.; Bardin, J.C.; Barends, R.; Biswas,
  R.; Boixo, S.; Brandao, F.G.S.L.; Buell, D.A.;  et~al.
\newblock Quantum supremacy using a programmable superconducting processor.
\newblock {\em Nature} {\bf 2019}, {\em 574},~505--510.
\newblock {\url{https://doi.org/10.1038/s41586-019-1666-5}}.

\bibitem[Zhong \em{et~al.}(2020)Zhong, Wang, Deng, Chen, Peng, Luo, Qin, Wu,
  Ding, Hu, Hu, Yang, Zhang, Li, Li, Jiang, Gan, Yang, You, Wang, Li, Liu, Lu,
  and Pan]{Pan2020}
Zhong, H.S.; Wang, H.; Deng, Y.H.; Chen, M.C.; Peng, L.C.; Luo, Y.H.; Qin, J.;
  Wu, D.; Ding, X.; Hu, Y.;  et~al.
\newblock Quantum computational advantage using photons.
\newblock {\em Science} {\bf 2020}, {\em 370},~1460--1463.
\newblock {\url{https://doi.org/10.1126/science.abe8770}}.

\bibitem[Daley \em{et~al.}(2022)Daley, Bloch, Kokail, Flannigan, Pearson,
  Troyer, and Zoller]{Zoller2022}
Daley, A.J.; Bloch, I.; Kokail, C.; Flannigan, S.; Pearson, N.; Troyer, M.;
  Zoller, P.
\newblock Practical quantum advantage in quantum simulation.
\newblock {\em Nature} {\bf 2022}, {\em 607},~667--676.
\newblock {\url{https://doi.org/10.1038/s41586-022-04940-6}}.

\bibitem[Liu \em{et~al.}(2021)Liu, Liu, Li, Fu, Yang, Song, Zhao, Wang, Peng,
  Chen, Guo, Huang, Wu, and Chen]{Liu2021}
Liu, Y.A.; Liu, X.L.; Li, F.N.; Fu, H.; Yang, Y.; Song, J.; Zhao, P.; Wang, Z.;
  Peng, D.; Chen, H.;  et~al.
\newblock Closing the "Quantum Supremacy" Gap: Achieving Real-Time Simulation
  of a Random Quantum Circuit Using a New Sunway Supercomputer.
\newblock In Proceedings of the Proceedings of the International Conference for
  High Performance Computing, Networking, Storage and Analysis; Association for
  Computing Machinery: New York, NY, USA,  2021; SC '21.
\newblock {\url{https://doi.org/10.1145/3458817.3487399}}.

\bibitem[Bulmer \em{et~al.}(2022)Bulmer, Bell, Chadwick, Jones, Moise, Rigazzi,
  Thorbecke, Haus, Vaerenbergh, Patel, Walmsley, and Laing]{Brumer2022}
Bulmer, J.F.F.; Bell, B.A.; Chadwick, R.S.; Jones, A.E.; Moise, D.; Rigazzi,
  A.; Thorbecke, J.; Haus, U.U.; Vaerenbergh, T.V.; Patel, R.B.;  et~al.
\newblock The boundary for quantum advantage in Gaussian boson sampling.
\newblock {\em Science Advances} {\bf 2022}, {\em 8},~eabl9236,
  \href{http://xxx.lanl.gov/abs/https://www.science.org/doi/pdf/10.1126/sciadv.abl9236}{{\normalfont
  [https://www.science.org/doi/pdf/10.1126/sciadv.abl9236]}}.
\newblock {\url{https://doi.org/10.1126/sciadv.abl9236}}.

\bibitem[Zhou \em{et~al.}(2020)Zhou, Stoudenmire, and Waintal]{Waintal2020}
Zhou, Y.; Stoudenmire, E.M.; Waintal, X.
\newblock What Limits the Simulation of Quantum Computers?
\newblock {\em Phys. Rev. X} {\bf 2020}, {\em 10},~041038.
\newblock {\url{https://doi.org/10.1103/PhysRevX.10.041038}}.

\bibitem[Cross \em{et~al.}(2019)Cross, Bishop, Sheldon, Nation, and
  Gambetta]{Gambetta19}
Cross, A.W.; Bishop, L.S.; Sheldon, S.; Nation, P.D.; Gambetta, J.M.
\newblock Validating quantum computers using randomized model circuits.
\newblock {\em Phys. Rev. A} {\bf 2019}, {\em 100},~032328.
\newblock {\url{https://doi.org/10.1103/PhysRevA.100.032328}}.

\bibitem[Pizzamiglio \em{et~al.}(2021)Pizzamiglio, Chang, Bondani, Montangero,
  Gerace, and Benenti]{Benenti21}
Pizzamiglio, A.; Chang, S.Y.; Bondani, M.; Montangero, S.; Gerace, D.; Benenti,
  G.
\newblock Dynamical Localization Simulated on Actual Quantum Hardware.
\newblock {\em Entropy} {\bf 2021}, {\em 23}.
\newblock {\url{https://doi.org/10.3390/e23060654}}.

\bibitem[Plenio and Virmani(2007)]{Plenio2007}
Plenio, M.B.; Virmani, S.
\newblock An Introduction to Entanglement Measures.
\newblock {\em Quantum Info. Comput.} {\bf 2007}, {\em 7},~1–51.

\bibitem[Horodecki \em{et~al.}(2009)Horodecki, Horodecki, Horodecki, and
  Horodecki]{Horodecki2009}
Horodecki, R.; Horodecki, P.; Horodecki, M.; Horodecki, K.
\newblock Quantum entanglement.
\newblock {\em Rev. Mod. Phys.} {\bf 2009}, {\em 81},~865--942.
\newblock {\url{https://doi.org/10.1103/RevModPhys.81.865}}.

\bibitem[Nielsen and Chuang(2000)]{nielsen00}
Nielsen, M.A.; Chuang, I.L.
\newblock {\em Quantum Computation and Quantum Information}; Cambridge
  University Press,  2000.

\bibitem[Facchi \em{et~al.}(2006)Facchi, Florio, and Pascazio]{Facchi2006}
Facchi, P.; Florio, G.; Pascazio, S.
\newblock Probability-density-function characterization of multipartite
  entanglement.
\newblock {\em Phys. Rev. A} {\bf 2006}, {\em 74},~042331.
\newblock {\url{https://doi.org/10.1103/PhysRevA.74.042331}}.

\bibitem[Brydges \em{et~al.}(2019)Brydges, Elben, Jurcevic, Vermersch, Maier,
  Lanyon, Zoller, Blatt, and Roos]{Brydges2019}
Brydges, T.; Elben, A.; Jurcevic, P.; Vermersch, B.; Maier, C.; Lanyon, B.P.;
  Zoller, P.; Blatt, R.; Roos, C.F.
\newblock Probing R\'enyi entanglement entropy via randomized measurements.
\newblock {\em Science} {\bf 2019}, {\em 364},~260--263.
\newblock {\url{https://doi.org/10.1126/science.aau4963}}.

\bibitem[Brand\~ao \em{et~al.}(2021)Brand\~ao, Chemissany, Hunter-Jones, Kueng,
  and Preskill]{brandao2021}
Brand\~ao, F.G.; Chemissany, W.; Hunter-Jones, N.; Kueng, R.; Preskill, J.
\newblock Models of Quantum Complexity Growth.
\newblock {\em PRX Quantum} {\bf 2021}, {\em 2},~030316.
\newblock {\url{https://doi.org/10.1103/PRXQuantum.2.030316}}.

\bibitem[Hayden and Preskill(2007)]{Hayden2007}
Hayden, P.; Preskill, J.
\newblock Black holes as mirrors: quantum information in random subsystems.
\newblock {\em Journal of High Energy Physics} {\bf 2007}, {\em 2007},~120.
\newblock {\url{https://doi.org/10.1088/1126-6708/2007/09/120}}.

\bibitem[Choi \em{et~al.}(2023)Choi, Shaw, Madjarov, Xie, Finkelstein, Covey,
  Cotler, Mark, Huang, Kale, et~al.]{choi2023}
Choi, J.; Shaw, A.L.; Madjarov, I.S.; Xie, X.; Finkelstein, R.; Covey, J.P.;
  Cotler, J.S.; Mark, D.K.; Huang, H.Y.; Kale, A.;  et~al.
\newblock Preparing random states and benchmarking with many-body quantum
  chaos.
\newblock {\em Nature} {\bf 2023}, {\em 613},~468--473.
\newblock {\url{https://doi.org/10.1038/s41586-022-05442-1}}.

\bibitem[Vatan and Williams(2004)]{vatan2004optimal}
Vatan, F.; Williams, C.
\newblock Optimal quantum circuits for general two-qubit gates.
\newblock {\em Phys. Rev. A} {\bf 2004}, {\em 69},~032315.
\newblock {\url{https://doi.org/10.1103/PhysRevA.69.032315}}.

\bibitem[Giraud(2007)]{Giraud2007}
Giraud, O.
\newblock Distribution of bipartite entanglement for random pure states.
\newblock {\em Journal of Physics A: Mathematical and Theoretical} {\bf 2007},
  {\em 40},~2793.
\newblock {\url{https://doi.org/10.1088/1751-8113/40/11/014}}.

\bibitem[Lloyd and Pagels(1988)]{Lloyd88}
Lloyd, S.; Pagels, H.
\newblock Complexity as thermodynamic depth.
\newblock {\em Annals of Physics} {\bf 1988}, {\em 188},~186--213.
\newblock {\url{https://doi.org/https://doi.org/10.1016/0003-4916(88)90094-2}}.

\bibitem[Emerson \em{et~al.}(2003)Emerson, Weinstein, Saraceno, Lloyd, and
  Cory]{Emerson03}
Emerson, J.; Weinstein, Y.S.; Saraceno, M.; Lloyd, S.; Cory, D.G.
\newblock Pseudo-Random Unitary Operators for Quantum Information Processing.
\newblock {\em Science} {\bf 2003}, {\em 302},~2098--2100,
  \href{http://xxx.lanl.gov/abs/https://www.science.org/doi/pdf/10.1126/science.1090790}{{\normalfont
  [https://www.science.org/doi/pdf/10.1126/science.1090790]}}.
\newblock {\url{https://doi.org/10.1126/science.1090790}}.

\bibitem[Emerson \em{et~al.}(2005)Emerson, Livine, and Lloyd]{Emerson2005}
Emerson, J.; Livine, E.; Lloyd, S.
\newblock Convergence conditions for random quantum circuits.
\newblock {\em Phys. Rev. A} {\bf 2005}, {\em 72},~060302.
\newblock {\url{https://doi.org/10.1103/PhysRevA.72.060302}}.

\bibitem[Weinstein and Hellberg(2005)]{Weinstein2005}
Weinstein, Y.S.; Hellberg, C.S.
\newblock Entanglement Generation of Nearly Random Operators.
\newblock {\em Phys. Rev. Lett.} {\bf 2005}, {\em 95},~030501.
\newblock {\url{https://doi.org/10.1103/PhysRevLett.95.030501}}.

\bibitem[Dahlsten \em{et~al.}(2007)Dahlsten, Oliveira, and Plenio]{Dahlsten07}
Dahlsten, O.C.O.; Oliveira, R.; Plenio, M.B.
\newblock The emergence of typical entanglement in two-party random processes.
\newblock {\em Journal of Physics A: Mathematical and Theoretical} {\bf 2007},
  {\em 40},~8081.
\newblock {\url{https://doi.org/10.1088/1751-8113/40/28/S16}}.

\bibitem[Oliveira \em{et~al.}(2007)Oliveira, Dahlsten, and Plenio]{Oliveira07}
Oliveira, R.; Dahlsten, O.C.O.; Plenio, M.B.
\newblock Generic Entanglement Can Be Generated Efficiently.
\newblock {\em Phys. Rev. Lett.} {\bf 2007}, {\em 98},~130502.
\newblock {\url{https://doi.org/10.1103/PhysRevLett.98.130502}}.

\bibitem[Cross \em{et~al.}(2019)Cross, Bishop, Sheldon, Nation, and
  Gambetta]{Cross19}
Cross, A.W.; Bishop, L.S.; Sheldon, S.; Nation, P.D.; Gambetta, J.M.
\newblock Validating quantum computers using randomized model circuits.
\newblock {\em Phys. Rev. A} {\bf 2019}, {\em 100},~032328.
\newblock {\url{https://doi.org/10.1103/PhysRevA.100.032328}}.

\bibitem[\ifmmode \check{Z}\else \v{Z}\fi{}nidari\ifmmode~\check{c}\else
  \v{c}\fi{}(2007)]{Znidaric07}
\ifmmode \check{Z}\else \v{Z}\fi{}nidari\ifmmode~\check{c}\else \v{c}\fi{}, M.
\newblock Optimal two-qubit gate for generation of random bipartite
  entanglement.
\newblock {\em Phys. Rev. A} {\bf 2007}, {\em 76},~012318.
\newblock {\url{https://doi.org/10.1103/PhysRevA.76.012318}}.

\bibitem[Pozniak \em{et~al.}(1998)Pozniak, Zyczkowski, and Kus]{Pozniak98}
Pozniak, M.; Zyczkowski, K.; Kus, M.
\newblock Composed ensembles of random unitary matrices.
\newblock {\em Journal of Physics A: Mathematical and General} {\bf 1998}, {\em
  31},~1059.
\newblock {\url{https://doi.org/10.1088/0305-4470/31/3/016}}.

\bibitem[Pelofske \em{et~al.}(2022)Pelofske, Bärtschi, and
  Eidenbenz]{Pelofske2022}
Pelofske, E.; Bärtschi, A.; Eidenbenz, S.
\newblock Quantum Volume in Practice: What Users Can Expect From NISQ Devices.
\newblock {\em IEEE Transactions on Quantum Engineering} {\bf 2022}, {\em
  3},~1--19.
\newblock {\url{https://doi.org/10.1109/TQE.2022.3184764}}.

\bibitem[Humphreys(2012)]{humphreys2012}
Humphreys, J.E.
\newblock {\em Introduction to Lie algebras and representation theory}; Vol.~9,
  Springer Science \& Business Media,  2012.

\bibitem[Bengtsson and Życzkowski(2017)]{bengtsson_życzkowski_2017}
Bengtsson, I.; Życzkowski, K.
\newblock {\em Geometry of Quantum States: An Introduction to Quantum
  Entanglement}, 2 ed.; Cambridge University Press,  2017.
\newblock {\url{https://doi.org/10.1017/9781139207010}}.

\bibitem[Giraud \em{et~al.}(2009)Giraud, \ifmmode \check{Z}\else
  \v{Z}\fi{}nidari\ifmmode~\check{c}\else \v{c}\fi{}, and Georgeot]{Giraud2009}
Giraud, O.; \ifmmode \check{Z}\else \v{Z}\fi{}nidari\ifmmode~\check{c}\else
  \v{c}\fi{}, M.; Georgeot, B.
\newblock Quantum circuit for three-qubit random states.
\newblock {\em Phys. Rev. A} {\bf 2009}, {\em 80},~042309.
\newblock {\url{https://doi.org/10.1103/PhysRevA.80.042309}}.

\bibitem[Dahlsten \em{et~al.}(2014)Dahlsten, Lupo, Mancini, and
  Serafini]{Dahlsten2014}
Dahlsten, O.C.O.; Lupo, C.; Mancini, S.; Serafini, A.
\newblock Entanglement typicality.
\newblock {\em Journal of Physics A: Mathematical and Theoretical} {\bf 2014},
  {\em 47},~363001.
\newblock {\url{https://doi.org/10.1088/1751-8113/47/36/363001}}.

\bibitem[Facchi \em{et~al.}(2008)Facchi, Florio, Parisi, and
  Pascazio]{Facchi2008}
Facchi, P.; Florio, G.; Parisi, G.; Pascazio, S.
\newblock Maximally multipartite entangled states.
\newblock {\em Phys. Rev. A} {\bf 2008}, {\em 77},~060304.
\newblock {\url{https://doi.org/10.1103/PhysRevA.77.060304}}.

\bibitem[McClean \em{et~al.}(2016)McClean, Romero, Babbush, and
  Aspuru-Guzik]{McClean2016}
McClean, J.R.; Romero, J.; Babbush, R.; Aspuru-Guzik, A.
\newblock The theory of variational hybrid quantum-classical algorithms.
\newblock {\em New Journal of Physics} {\bf 2016}, {\em 18},~023023.
\newblock {\url{https://doi.org/10.1088/1367-2630/18/2/023023}}.

\bibitem[Moll \em{et~al.}(2018)Moll, Barkoutsos, Bishop, Chow, Cross, Egger,
  Filipp, Fuhrer, Gambetta, Ganzhorn, Kandala, Mezzacapo, Müller, Riess,
  Salis, Smolin, Tavernelli, and Temme]{Moll2018}
Moll, N.; Barkoutsos, P.; Bishop, L.S.; Chow, J.M.; Cross, A.; Egger, D.J.;
  Filipp, S.; Fuhrer, A.; Gambetta, J.M.; Ganzhorn, M.;  et~al.
\newblock Quantum optimization using variational algorithms on near-term
  quantum devices.
\newblock {\em Quantum Science and Technology} {\bf 2018}, {\em 3},~030503.
\newblock {\url{https://doi.org/10.1088/2058-9565/aab822}}.

\bibitem[Cerezo \em{et~al.}(2021)Cerezo, Arrasmith, Babbush, Benjamin, Endo,
  Fujii, McClean, Mitarai, Yuan, Cincio, and Coles]{Cerezo2021}
Cerezo, M.; Arrasmith, A.; Babbush, R.; Benjamin, S.C.; Endo, S.; Fujii, K.;
  McClean, J.R.; Mitarai, K.; Yuan, X.; Cincio, L.;  et~al.
\newblock Variational quantum algorithms.
\newblock {\em Nature Reviews Physics} {\bf 2021}, {\em 3},~625--644.
\newblock {\url{https://doi.org/10.1038/s42254-021-00348-9}}.

\bibitem[Khaneja and Glaser(2001)]{khaneja2001cartan}
Khaneja, N.; Glaser, S.J.
\newblock Cartan decomposition of SU (2n) and control of spin systems.
\newblock {\em Chemical Physics} {\bf 2001}, {\em 267},~11--23.
\newblock {\url{https://doi.org/10.1016/S0301-0104(01)00318-4}}.

\bibitem[Khaneja \em{et~al.}(2001)Khaneja, Brockett, and
  Glaser]{khaneja2001time}
Khaneja, N.; Brockett, R.; Glaser, S.J.
\newblock Time optimal control in spin systems.
\newblock {\em Phys. Rev. A} {\bf 2001}, {\em 63},~032308.
\newblock {\url{https://doi.org/10.1103/PhysRevA.63.032308}}.

\end{thebibliography}

\end{adjustwidth}
\end{document}